\documentclass[pre,amsmath,showpacs]{revtex4}
\usepackage{graphicx}
\usepackage{color}
\usepackage{bm}
\usepackage{epsfig}
\usepackage{latexsym}
\usepackage{pifont}
\usepackage{float}
\usepackage[utf8]{inputenc}
\graphicspath{ {figure/} }
\usepackage{siunitx}

\begin{document}

\newcommand{\be}{\begin{equation}}
\newcommand{\ee}{\end{equation}}
\newcommand{\bea}{\begin{eqnarray}}
\newcommand{\eea}{\end{eqnarray}}
\newcommand{\nn}{\nonumber}

\today

\title{Actin filaments growing against an elastic membrane: Effect of membrane tension}
\author{Raj Kumar Sadhu and Sakuntala Chatterjee}
\affiliation{Department of Theoretical Sciences, S. N. Bose National Centre for Basic Sciences, Block  JD, Sector  III, Salt Lake, Kolkata  700106, India. }

\pacs {05.40.-a, 87.16.aj, 87.16.Ka, 87.15.A-}

\begin{abstract}
We study the force generation by a set of parallel actin filaments growing against an elastic membrane. The elastic membrane tries to stay flat and any deformation from this flat state, either caused by thermal fluctuations or due to protrusive polymerization force exerted by the filaments, costs energy. We study two lattice models to describe the membrane dynamics. In one case, the energy cost is assumed to be proportional to the absolute magnitude of the height gradient (gradient model) and in the other case it is proportional to the square of the height gradient (Gaussian model). For the gradient model we find that the membrane velocity is a non-monotonic function of the elastic constant $\mu$, and reaches a peak at $\mu=\mu^\ast$. For $\mu < \mu^\ast$ the system fails to reach a steady state and the membrane energy keeps increasing with time. For the Gaussian model, the system always reaches a steady state and the membrane velocity decreases monotonically with the elastic constant $\nu$ for all nonzero values of $\nu$. Multiple filaments give rise to protrusions at different regions of the membrane and the elasticity of the membrane induces an effective attraction between the two protrusions in the Gaussian model which causes the protrusions to merge and a single wide protrusion is present in the system. In both the models, the relative time-scale between the membrane and filament dynamics plays an important role in deciding whether the shape of elasticity-velocity curve is concave or convex. Our numerical simulations agree reasonably well with our analytical calculations.  
\end{abstract}
\maketitle

\section{Introduction}
 Cell motility plays an important role in many important biological processes such as morphogenesis, wound repair, cancer invasion etc. \cite{review1, review2, review3, review4, review5}. Actins and microtubules are cytoskeletal proteins whose polymerization and depolymerization  generate significant force that induce the cell motility. Inside the cell, these filaments grow against the cell membrane which acts as a biological barrier. The polymerization force generated by these filaments is measured in many \textit{in vitro} cases by applying an opposing external load on the barrier. The velocity of the barrier decreases with the external load and this dependency, known as the force-velocity curve, is an important characteristic of the force generation mechanism.

The plasma membrane against which the actin filaments grow has a role of central importance in the motility process \cite{review6, review7, review8, review9}. The membrane tension couples different processes that are part of cell migration  \cite{NilsGauthier2011} and hence it is important to understand how the membrane tension affects actin polymerization. Extension of the plasma membrane that takes place during cell motility and cell spreading depends strongly on membrane tension. In Ref. \cite{DRaucher2000} it was shown that when the membrane tension is decreased by adding certain ampiphillic compounds, the lamellopodial extension rate increases, while in situations where membrane tension is increased using osmotic methods, the extension rate drops. Although the membrane tension is generally considered as an obstacle to cell movement, it was shown in  Ref. \cite{EllenLBatchelder2011} that the membrane tension also optimizes the motility in {\sl Caenorhabditis elegans} sperm cells by streamlining actin polymerization in the direction of movement. In an {\it in vitro} reconstitution of lipid bilayer, it has been found that membrane elasticity causes formation of filament bundles that support membrane protrusions from branched filament network \cite{AllenPLiu2008}. A flexible membrane enhances formation of filopodial protrusion and also allows merging of smaller neighboring protrusions into a larger one \cite{ErdincAtilgan2006}.

Above studies show that the elastic properties of a flexible membrane have significant influence on actin filament polymerization and force generation process. This gives rise to a more general question: What happens when a flexible, elastically deformable obstacle is placed in the path of growing filaments? Some interesting results were obtained in this direction in Ref. \cite{baumgaertner2010}, where an elastic membrane was described using a solid-on-solid model \cite{lipowsky94, AVolmer1998} and the leading edge of the actin mesh was modelled by an advancing uncorrelated front. In absence of the advancing mesh, the equilibrium thermal fluctuations of the membrane can be characterized by Edwards-Wilkinson universality class \cite{ew,baumgaertner2010}. However, when driven by the polymerizing mesh, the growth exponents of the membrane shape fluctuations show a slow crossover to Kardar-Parisi-Zhang universality class \cite{kpz,baumgaertner2010}. This similar modeling strategy was generalized in Ref. \cite{baumgaertner2012} to understand certain  aspects of crawling motion of a cell on a flat substrate, particularly, how the asymmetry between the leading and trailing edge of the membrane decides the shape and velocity of the membrane.

In this paper, we study a coupled system consisting of an elastic membrane and a set of growing filaments, which pushes the membrane from below. Due to elastic energy, the membrane tries to stay flat and any deformation from this flat state costs energy. A height field describes the membrane configuration such that in a flat state, height is same everywhere on the membrane and whenever a height gradient develops, it costs energy.  We consider two different lattice models to describe the membrane dynamics. In one case, we use the model introduced in Ref. \cite{lipowsky94} to describe protrusions in a membrane and studied later in Ref. \cite{baumgaertner2010}. Here, the energy cost of a local membrane deformation is assumed to be proportional to the absolute magnitude of the local height gradient. Below we refer this model as the ``gradient model". In the second case, the energy cost is proportional to the square of the local height gradient \cite{AVolmer1998}. We call it the ``Gaussian model". The filaments are modelled by rigid parallel rods, which are pushing against the membrane and causing its deformation. We are interested to find out how the membrane elasticity affects the growth of the filaments and how the polymerization force exerted by the filaments affects the membrane deformations.

We find that for the gradient model, the average membrane velocity shows a peak as a function of the membrane tension $\mu$. This is somewhat surprising since a larger membrane tension is expected to make the polymerization process more costly and hence the growth process slower. We show that the peak results from the competition between the polymerization force of the filaments and the elastic force of the membrane. We also show that for small $\mu$, the system does not have a steady state and the total contour length of the membrane keeps increasing with time. For large values of $\mu$, the system reaches a steady state and the membrane velocity then decreases with $\mu$. For the Gaussian model, however, the membrane velocity decreases monotonically with the membrane elastic constant $\nu$ and steady state is reached for all nonzero values of $\nu$. In both the models, the elasticity-velocity curve plays similar role as the force-velocity characteristic obtained in earlier studies \cite{marcy2004,baudry2011,baudry2014,theriot2005,peskin1993,Sadhu2016,hansda2014,
carlsson2014}. We show that the relative time-scale between the membrane dynamics and the filament dynamics is crucial to determine the qualitative shape of the elasticity-velocity curve. The importance of the relative time scale was also highlighted in our earlier work \cite{Sadhu2016} where instead of an elastically deformable membrane, we had used a Kardar-Parisi-Zhang surface whose height fluctuations were biased in the direction opposite to that of polymerization. Even in that case, we had shown that the force-velocity curve can change its shape depending on the relative time scale between the filament and surface dynamics \cite{Sadhu2016}. These recent results, along with the current ones reported here, emphasize the importance of taking into account the independent (thermal) fluctuations of the membrane shape and position, while studying force generation by growing filaments. In many recent modelling approaches this aspect has been ignored and a rigid barrier with no shape fluctuations has been considered instead \cite{kirone, ddas2014, hansda2014}.

This paper is organized as follows. In Sec. \ref{sec:model}, we describe our models. Our results for the gradient model are presented in Sec. \ref{sec:grad} and the results for the Gaussian model are described in Sec. \ref{sec:gaussian} and conclusions are in Sec. \ref{sec:sum}. 

\section{Description of the model}
\label{sec:model}
\begin{figure}[h!]
\includegraphics[scale=1.4]{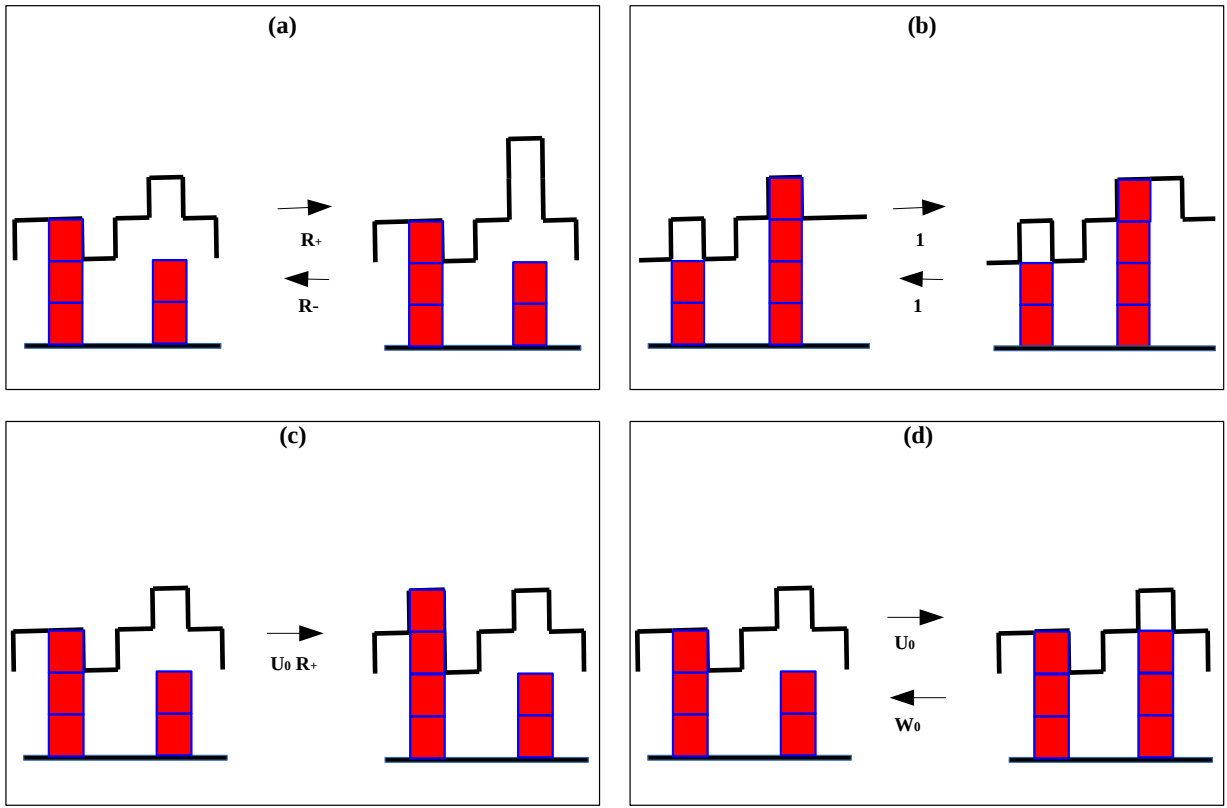}
\caption{Schematic representation of our model. The square blocks show actin monomers, which join together to form rod-like filaments. The thick solid line represents the shape of the elastic membrane. \textbf{(a):} A bulk site of the membrane thermally fluctuates and changes its height by an amount $d$, which in turn changes the membrane contour length by $2d$. The forward process increases the energy and occur with rate $R_+$ while the reverse process decreases the energy and happens with rate $R_-$. \textbf{(b):} A bulk site of the membrane changes its height by an amount $d$ but the energy remains same and thus the movement happens with rate unity. \textbf{(c):} A bound filament pushes the binding site by an amount $d$ that costs energy and occurs with rate $U_0 R_+$. \textbf{(d):} A free filament polymerizes (depolymerizes) with rate $U_0$ ($W_0$).}
\label{model}
\end{figure}

Our system consists of a set of $N$  parallel filaments growing against an elastic membrane as shown in Fig. \ref{model}. The membrane is modelled as a one dimensional lattice of length $L$ and lattice constant $d$. At each site $i$ of the lattice a height $h_i$ is assigned. For a completely flat membrane, when height of all sites are the same, the elastic energy is minimum. Presence of a local height gradient stretches the membrane and costs energy. In the absence of any filament, the membrane undergoes equilibrium thermal fluctuations and the probability to obtain a particular height profile $\{h_i \}$ follows Boltzmann measure. 

We consider two different models to describe the equilibrium membrane dynamics. In the first case, the membrane is modelled as a solid-on-solid (SOS) surface without diffusion \cite{nelson} and the Hamiltonian has the form \cite{lipowsky94, AVolmer1998, baumgaertner2010}
\be
{\cal H}_1=\mu \sum_{i=1}^{L}|h_i-h_{i+1}|
\label{eq:hamiltonian} 
\ee
where $\mu$ is the tension. The sum in Eq. (\ref{eq:hamiltonian}) is related to the total contour length $\cal C$ of the particular height configuration of the membrane, such that ${\cal C} =\sum_{i=1}^{L}|h_i-h_{i+1}| + Ld$. Note that since the energy is linear in $\cal C$, the energy cost for increasing the contour length by an amount $\delta$ just depends on $\delta$ and is independent of $\cal C$. We call Eq. (\ref{eq:hamiltonian}) the ``gradient model". This model was used in Ref. \cite{lipowsky94} to study thermally excited protrusions in lipid membranes. In Ref. \cite{AVolmer1998}, the critical behavior of this model was analysed using functional renormalization. In Ref. \cite{baumgaertner2010}, Eq. (\ref{eq:hamiltonian}) was used to describe a membrane and scaling properties of the height fluctuations were studied when the membrane was driven by an advancing uncorrelated front representing the actin meshwork. Towards the end of the paper, we present a discussion on this.

In our second model, the membrane is described as a Gaussian surface with the Hamiltonian \cite{AVolmer1998, baumgaertner2012}
\be
{\cal H}_2=\nu \sum_{i=1}^{L}(h_i-h_{i+1})^2
\label{eq:hgauss} 
\ee
where $\nu$ denotes the elastic constant. In this case, energy depends on the  square of the local height gradients, and hence the energy cost to extend the membrane by a certain amount also depends on the contour length $\cal C$. For large $\cal C$ values, the energy cost also goes up, as expected for an elastic body. We refer this model as `Gaussian model' below. To keep our description simple, we have neglected the bending energy of the membrane and have only considered its elastic energy here. We assume periodic boundary condition on the membrane, $h_{L+1}=h_1$.

The membrane can undergo independent thermal fluctuations in its local height [see Figs. \ref{model}(a) and \ref{model}(b)] and tends to minimize its elastic energy. In our lattice model, we assume that as a result of these fluctuations, the local height can increase or  decrease by a discrete amount $\delta$. The rates of the dynamical moves that changes the energy by an amount $\triangle E$ are assumed to satisfy local detailed balance 
\be
\frac{R_+}{R_-}=e^{-\beta  \Delta E}
\label{eq:ldb}
\ee 
where $R_+$ ($R_-$) is the rate of those processes that increases (decreases) the energy by an amount $\Delta E$ [Fig. \ref{model}a]. For $\Delta E =0$, the rate is taken to be unity [Fig. \ref{model}b].

The filaments are modelled as rigid polymers, composed of few rodlike monomers of length $d$. A (de)polymerization event increases (decreases) the length of a filament by an amount $d$. Throughout this work, we consider $\delta$, the unit of barrier height fluctuation, and the monomer size $d$ to be the same. There are two types of filaments we need to consider: free filaments, which are not in touch with the membrane, and bound filaments, whose tip is in contact with a membrane site. The point of contact is called the binding site. For a free filament, the polymerization process happens with a rate $U_0$ and depolymerization happens with a rate $W_0$ [see Fig. \ref{model}(d)]. However, for a  bound filament, a polymerization process increases the height of the binding site by an amount $d$ [see Fig. \ref{model}(c)]  and hence an energy cost $\Delta E$ is involved. Note that the change in energy can be positive or negative, or even zero, depending on the local height configuration around the binding site. For a positive (negative) energy cost, the bound filament polymerization rate is taken to be $U_0 R_+$  ($U_0 R_-$), while for zero energy cost, the rate is simply $U_0$. The depolymerization rate of bound filament does not involve any membrane movement and hence is equal to $W_0$.

The elastic interaction of the membrane tends to keep the membrane flat and the bound filaments generally grow by causing protrusions in the membrane. Although in principle, it is possible that the local height configuration around the binding site is such that the polymerization of bound filament actually releases some elastic energy, such configurations are rare and most of the time elastic force acts against polymerization force. The elasticity-velocity curve therefore plays a similar role as the force-velocity curve measured in many earlier studies \cite{marcy2004,baudry2011,baudry2014,theriot2005,peskin1993,Sadhu2016,hansda2014,
carlsson2014}. 

Note that the height of the membrane at the binding sites should be such that the membrane always stays above the filament tips, and this puts some restrictions in height fluctuations at the binding sites. Any height fluctuation that brings a binding site at a lower height than the filament tip is forbidden. Everywhere else in the membrane, the height fluctuations will occur in accordance with Eq. (\ref{eq:ldb}).

We perform simulations using kinetic Monte Carlo technique. The relative time scale between the filament dynamics and the membrane dynamics is quantified by a parameter 
${\cal S}$. For a system consisting of $N$ filaments and $L$ membrane sites, each Monte Carlo time-step consists of $N$ filament updates and ${\cal S}$ independent membrane updates. In case of a bound filament polymerization, the membrane height is also simultaneously updated at the binding site. More specifically, for $N < {\cal S}$, we first choose a filament at random and perform polymerization or depolymerization move as described above. Then we choose ${\cal S}/N$ membrane sites in random sequential order and update them (for noninteger values of ${\cal S}/N$, we replace them by the nearest integer). Repeating this process $N$ times completes one Monte Carlo step. Similarly, for $N > {\cal S}$, we first perform $N/{\cal S}$ filament updates and then choose one membrane site at random and update it; repeat this process ${\cal S}$ times and that defines one Monte Carlo step. Starting from an initially flat membrane and all filaments of length $d$ ({\sl i.e.}, each filament consists of only one monomer), the system undergoes time-evolution and after a large number of Monte Carlo steps, we perform our measurements. The value of parameters $U_0$ and $W_0$ are taken from Refs. \cite{review1,pollard} and the value of $d$ is taken from Refs. \cite{review1, hansda2014}. All these simulation parameters are listed in Table \ref{table} 
\begin{table}
\begin{center}
 \begin{tabular}{|c | c| c|} 
 \hline
 $U_0 $ & Free filament polymerization rate & $2.784 s^{-1}$\\
 \hline
 $W_0 $ & Filament depolymerization rate & $1.4 s^{-1}$\\ 
 \hline
 $d$ & Size of an actin monomer & $2.7 nm$ \\ 
 \hline
 $T$ & Temperature & $300 K$ \\ 
 \hline
\end{tabular}
 \caption{Parameters used in our simulation.}
 \label{table}
\end{center}
\end{table}

\section{Results for gradient model} \label{sec:grad}
In the gradient model, since the elastic energy of the membrane is proportional to its contour length, the change in local height by an amount $d$ that causes $\cal C$ to change by $2d$, brings about a change $2 \mu d$ in the energy. In our simulation, we choose $R_+=e^{-\beta \mu d}$ and $R_-=e^{\beta \mu d}$. The bound filament polymerization that leads to an increase (decrease) in energy happens with rate $U_0 R_+$ ($U_0 R_-$). All other movements of the membrane where energy does not change, occur with rate unity. We first present the results for a single filament and ${\cal S}/L =1$.

\subsection{Peak in the elasticity-velocity curve for single filament} 
\label{sec:peak}
\begin{figure}[h!]
\includegraphics[scale=1.4]{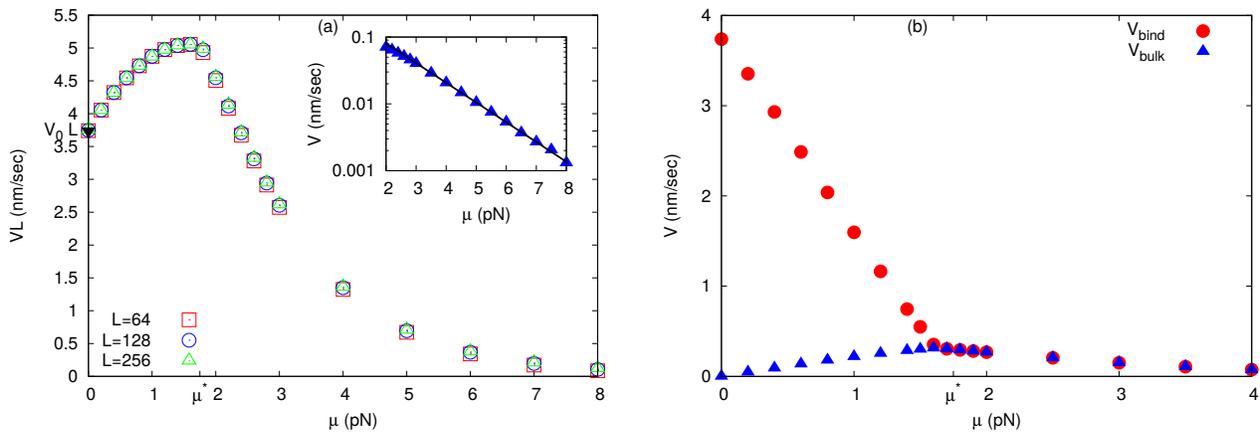}
\caption{$\mu-V$ curve for a single filament. \textbf{(a):} The average velocity of the membrane as a function of $\mu$ shows a peak at $\mu^*$. While $V \sim 1/L$, the peak-position $\mu^\ast$ does not depend on $L$. The solid triangle on the $y$-axis marks $V_0=\frac{d}{L}(U_0-W_0)$, which is the expected value of $V$ at $\mu=0$. Inset: For large $\mu$, the membrane velocity decreases exponentially with a decay constant $ \simeq 0.67$, which is close to the value of $\beta d$. Here, we have used  $L=64$. \textbf{(b):} The variation of average velocity of the binding site ($V_{bind}$) and the average velocity of a bulk site ($V_{bulk}$) with $\mu$,  shows that $V_{bind}$ decreases monotonically with $\mu$ while $V_{bulk}$ increases with $\mu$ for $\mu < \mu^\ast $. For $\mu > \mu^\ast$, these two velocities are equal. Here we have used $L=16$. For all the above plots, we have used ${\cal S}/L=1$. Other simulation parameters are as in Table \ref{table}.}
\label{mu_vs_v} \end{figure}

The polymerization of the bound filament pushes the membrane upward and gives rise to a nonzero membrane velocity, which is measured as the rate of change of the average membrane height in the long time limit. We measure the velocity $V$ as a function of $\mu$ and present our data in Fig. \ref{mu_vs_v}(a), for $N=1$ and ${\cal S}/L=1$. We find that $V$ shows a nonmonotonic dependence on $\mu$: Starting from a nonzero value for $\mu =0$, it increases with $\mu$ for small $\mu$ values, and after reaching a peak at a certain $\mu^\ast$, the velocity  decreases again. Presence of a peak is somewhat surprising, since increasing $\mu$ ought to make it more difficult for the filament to push against the membrane. We also find that $V$ scales as $1/L$. In the remaining part of this sub-section, we explain different aspects of this data in detail. 

First we consider $\mu=0$. In this case, there is no energy cost involved in stretching the membrane, and thus the single filament present in the system polymerizes with rate $U_0$, irrespective of whether it is free or bound to the membrane. The velocity of the barrier in this case turns out to be $V_0 = d(U_0-W_0)/L $. We briefly present the calculation here. Let $p_0$ be the probability that the filament is in the bound state. The height at the binding site can (a) increase due to bound filament polymerization that happens with an effective rate $U_0 p_0$, (b) increase due to thermal fluctuation that happens with rate $1$, or (c) decrease due to thermal fluctuation, provided the  filament is not in a bound state (since the membrane always needs to stay above the filament tip) and this process happens with effective rate $(1-p_0)$. It follows from here that the average velocity of the membrane can be written as $V_0 = V(\mu =0) = \dfrac{d}{L} p_0 (U_0+1) $, where the pre-factor $d/L$ is the change in average height of the membrane due to $d$-unit change in the binding site height. Here, we have assumed ${\cal S}/L=1$. The contact probability $p_0$ can be calculated by noting that the height difference between the filament tip and the binding site performs a biased random walk, with the restriction that it cannot cross the origin and become negative \cite{Sadhu2016}. Our simple calculation presented in Appendix \ref{app:p0} yields $p_0 = \dfrac{U_0-W_0}{1+U_0} $, which gives the required expression for $V_0$. Comparison with our simulation data in Fig. \ref{mu_vs_v}(a) show good agreement.

Next, we explain the presence of the peak in the $\mu-V$ curve. Note that for $\mu=0$ the system has no steady state and the filament pushes the binding site upward, while other $(L-1)$ sites of the system which are not coupled ($\mu$ being zero), undergo equilibrium fluctuations and show no net velocity. Consequently, the height difference between the binding site and the bulk sites and hence the contour length $\cal C$ keeps increasing with time. For nonzero $\mu$ there is an energy cost associated with stretching the membrane. This elastic force tries to reduce ${\cal C}$ and the bulk sites feel an upward pull towards the binding site. The strength of this pull increases as $\mu$ increases and this explains why the membrane velocity increases with $\mu$. However, for small values of $\mu$, this elastic force is not strong enough to counter the polymerization force exerted by the filaments, and the bulk sites are not able to catch up with the binding site, which still moves at a larger velocity and $\cal C$ keeps increasing with time. Finally, when $\mu$ reaches a critical value $\mu^\ast$, such that the elastic force exactly balances the polymerization force, the average velocity of the bulk sites becomes equal to that of the binding site.
As $\mu$ is increased further, the elastic force becomes stronger than the polymerization force and it becomes increasingly difficult for the filament to push the binding site. However, as soon as a successful polymerization takes place, and the binding site height increases, the bulk sites quickly catch up because the elastic energy associated with a nonflat profile is high for $\mu$ values in this range. The whole membrane now moves with the same velocity and $\cal C$ stabilizes. The membrane velocity in this case is dominated by the polymerization events at the binding site, the rate of which is $U_0 \exp(-\beta \mu d)$. Thus the membrane velocity decreases exponentially with $\mu$ in this range [see data in Fig. \ref{mu_vs_v}(a), inset].

\begin{figure}[h!]
\includegraphics[scale=1.4]{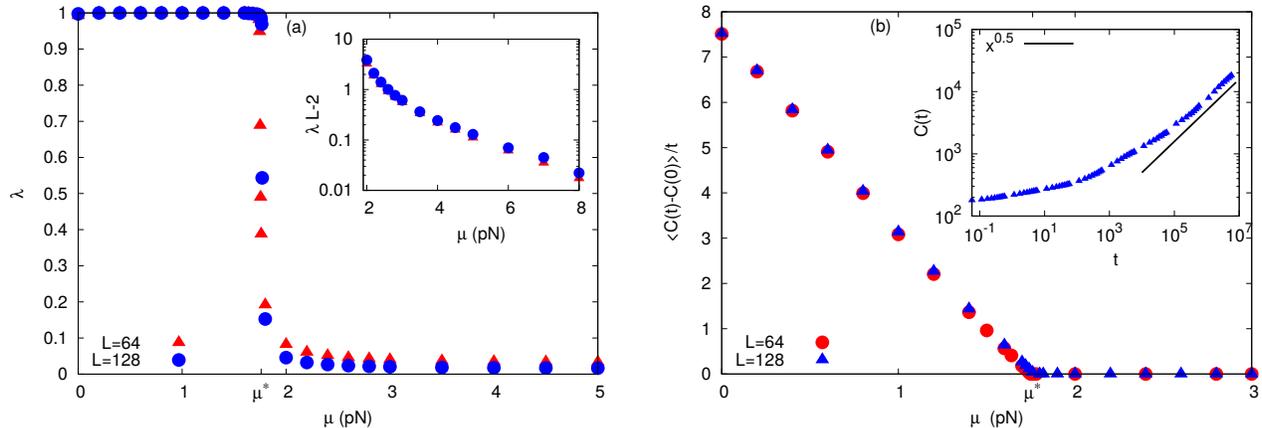}
\caption{ The membrane contour length keeps growing for $\mu < \mu^\ast$ and stabilizes for $\mu > \mu^\ast$. \textbf{(a):} The ratio $\lambda=1$ for $\mu <\mu^\ast$ and decreases sharply at $\mu^\ast$ to its asymptotic value $2/L$ for large $\mu$. Inset: The saturation of $\lambda$ at large $\mu$ happens exponentially. \textbf{(b):} The quantity $\frac{<{\cal C}(t) - {\cal C}(0)>}{t}$, for large $t$, decreases with $\mu$ and becomes zero above $\mu^*$. Inset:  Variation of contour length with time at $\mu^* = 1.748 pN$. We see that after large enough time, ${\cal C}(t)$ grows with time as $t^{0.5}$ with a diffusion constant $D=37.41 \pm 7.19 nm^2/s$.  The analytically calculated value of $D=41.352 nm^2/s$ which is close to the numerical value. Here we use $L=64$. For both panels ${\cal S}/L=1$. Other simulation parameters are as in Table \ref{table}.}
\label{fig:cl} \end{figure}

To verify the above mechanism, we measure the velocity of the binding site and the bulk sites separately and plot the data in  Fig. \ref{mu_vs_v}(b). As argued above, we find that for small $\mu$, the binding site velocity is higher than the bulk site and these two velocities become equal for $\mu \geq \mu^\ast$. The difference between the binding site and the bulk site velocities can be alternatively measured by $\frac{<{\cal C}(t) - {\cal C}(0)>}{t}$, for large t, and our plot in Fig. \ref{fig:cl}b shows that for $\mu < \mu^\ast$ this quantity decreases linearly with $\mu$ and becomes zero at $\mu^\ast$. We also measure the ratio $\lambda= <(|h_{b+1}-h_b| + |h_{b}-h_{b-1}| + 2d )/{\cal C}> $, averaged over different configurations. Here, $b$ denotes the binding site. $\lambda$ gives the fraction of contour length that is contained between the binding site and its two neighbors. For $\mu < \mu^{\ast}$, when the binding site moves faster than the rest of the membrane, for large times, this fraction is very close to 1, since the length of the rest of the membrane becomes negligible compared to the growing separation between the binding site and its neighbors. Our data in Fig. \ref{fig:cl}(a) show that $\lambda$ stays at $1$ for small $\mu$ and then shows a sharp fall at $\mu = \mu^\ast$. For $\mu \gg \mu^\ast$, when the polymerization is almost impossible and the membrane velocity is close to zero, the membrane is flat and $\lambda$ then becomes $2/L$. In this limit, the membrane behaves like a rigid barrier, which was studied in Refs. \cite{kirone, peskin1993, krawczyk2011, ddas2014, carlsson2014, hansda2014}. We find that after the sharp jump at $\mu = \mu^\ast$, the ratio of the contour lengths decreases exponentially with $\mu$ to the asymptotic value $2/L$  [Fig. \ref{fig:cl}(a), inset]. Note that both for small and large $\mu$, the binding site is the only site that is being pushed by the filament and the dynamics of bulk sites still follow local detailed balance. The overall velocity of the membrane is thus generated by the drive present at the single binding site and this is the reason $V$ scales inversely with the system size $L$. 

\subsection{Absence of steady state for $\mu < \mu^\ast$} 
For $\mu < \mu^\ast$, when the binding site and bulk site velocities are different, and ${\cal C}$ keeps increasing with time, the system does not have a steady state. This seems somewhat surprising and in this subsection we discuss this issue in detail. 

Let us define $z = 2h_b -h_{b+1}-h_{b-1}$. Our data in Appendix \ref{app:pz} [Fig. \ref{height_gradient}(a)] show that for $\mu < \mu^\ast$, the sign of $z$ is always positive and even for $\mu >  \mu^\ast$, the probability to find negative $z$ is negligible. This is expected, since for small $\mu$, the binding site always stays above its neighbors, and for large $\mu$, when the membrane is flat, the binding site is most of the time at the same level with its neighbors, but does not typically fall below that level. We show below that $z$ performs a biased random walker with a reflecting boundary condition at the origin. 

First note that $z$ can increase either due to increase of $h_b$ (from bound filament polymerization or thermal fluctuation at the binding site) which happens with rate $e^{-\beta \mu d} (p_0U_0+1)$, or due to decrease of $h_{b \pm 1}$, which happens with rate $1$. This is because our data in Appendix \ref{app:pz} (Fig. \ref{height_gradient}b) show that the sites $(b \pm 1)$ almost always have one neighbor (site $b \pm 2$) at a lower height, and another neighbor (site $b$) at a  higher height. From Eq. (\ref{eq:hamiltonian}) it then follows that the height at the sites $(b \pm 1)$ can increase or decrease without any energy cost and rate of such processes is taken to be  $1$ [see, for example, Fig. \ref{model}(b)]. The value of $z$ can decrease because of decrease in $h_b$ which releases elastic energy and happens with rate $e^{\beta \mu d}(1-p_0)$, or because of increase of $h_{b \pm 1}$, which again happens with rate unity, as explained above. Let us define $P(z,t)$ as the probability for $z$ at time $t$. Then the change in $P(z,t)$ in a small time $\triangle t$ can be written as,
\bea
P(z,t+\triangle t) - P(z,t)= \triangle t [(1-p_0) e^{\beta \mu d} P(z+2d,t) + 2P(z+d,t) + 2P(z-d,t) + (1+U_0p_0)e^{-\beta \mu d}P(z-2d,t) 
\nonumber \\
- \{(1-p_0) e^{\beta \mu d} + 4 + (1+U_0p_0)e^{-\beta \mu d}\} P(z,t)]
\label{eq:pz}
\eea
Here, we have considered $z >0$ (supported by data in Fig. \ref{height_gradient}a) and put a reflecting boundary at the origin $z=0$, such that $z$ can never be negative. In the continuum limit, Eq. (\ref{eq:pz}) becomes a Fokker-Planck equation for a biased random walker with drift and diffusion given by, respectively, $v= 2d\{(1+U_0p_0) e^{-\beta \mu d} -(1-p_0)e^{\beta \mu d} \}$ and $D= 2d^2 \{(1-p_0) e^{\beta \mu d} + (1+U_0p_0) e^{-\beta \mu d} +1 \}$.

It can be easily seen that for small $\mu$, the bias $v$ is positive while for large $\mu$, the bias $v$ is negative. At $\mu = \mu^*$, $v$ becomes zero and the system shows unbiased diffusion with diffusivity $D$. The value of $\mu^*$  can be obtained by equating $v$ to zero and using the expression for $p_0$ from Appendix \ref{app:p0} which gives the resulting equation
\be
W_0 e^{3 \beta \mu^\ast d} - U_0 e^{2 \beta \mu^\ast d} - (U_0 ^2 -U_0 W_0 + U_0) e^{\beta \mu^\ast d} + U_0 =0.
\ee
Numerical solution of the above equation gives only one physical solution for 
$\mu^\ast$. We find $\mu^* =1.81 pN$, which is close to the value $1.748 \pm 0.004 pN$ observed in simulations.

Due to the reflecting boundary condition at the origin, this implies that for $\mu < \mu^\ast$, the system does not reach a steady state, and $\langle z \rangle $ increases linearly with time, while for $\mu > \mu^\ast$, there exists a steady state and  $\langle z \rangle $ reaches a time-independent value. At $\mu=\mu^\ast$, we have an unbiased random walker with reflecting boundary at the origin, for which there is no steady state either, but $\langle z \rangle $ in this case grows diffusively with time with a diffusion constant that matches our analytical expression given above [see Fig. \ref{fig:cl}(b), inset].

\subsection{Faster membrane dynamics lowers $\mu^*$}
\begin{figure}[h!]
\includegraphics[scale=1.4]{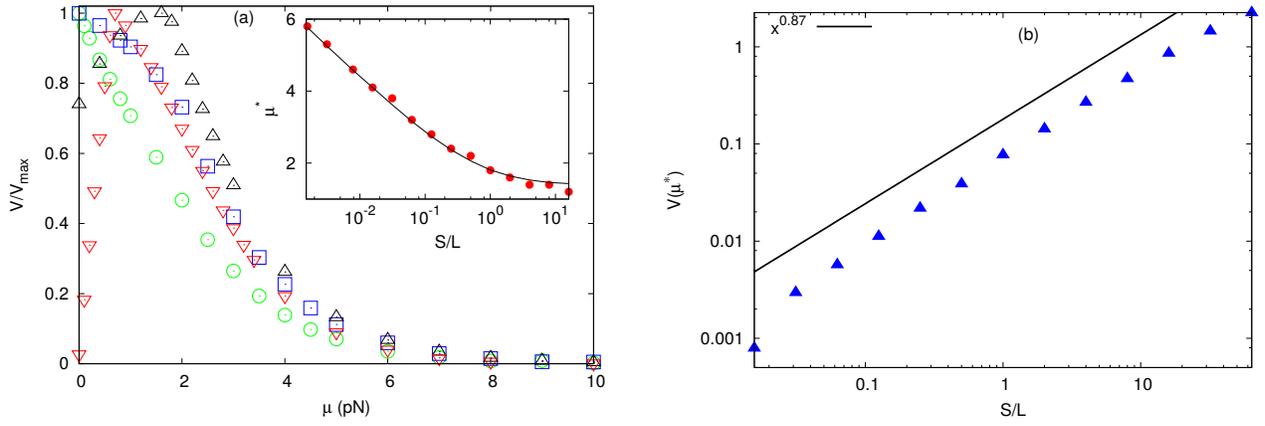}
\caption{Peak position and peak height of $\mu-V$ curve depends on ${\cal S}/L$. \textbf{(a):} Scaled $V$ as a function of $\mu$ for different ${\cal S}/L$ values. The scaling factor $V_{max}$ has been used such that the data for different ${\cal S}/L$ values can be compared. In this plot the symbol $\circ$ correspond to ${\cal S}/L=2^{-2}$ ($V_{max} \simeq \num{5.84e-2} nm/s$), the $\Box$ correspond to ${\cal S}/L=2^{-1}$ ($V_{max} \simeq \num{5.84e-2} nm/s$), the $\triangle$ correspond to ${\cal S}/L=1$ ($V_{max} \simeq \num{7.89e-2} nm/s$) and the $\nabla$ correspond to ${\cal S}/L=2^6$ ($V_{max} \simeq 2.23 nm/s$). For smaller values of ${\cal S}/L$, the curve is monotonic as the velocity is mostly determined by the binding site. As ${\cal S}/L$ increases, the $\mu-V$ curve develops a peak at $\mu^\ast$. Inset: $\mu^\ast$ decreases with ${\cal S}/L$. The discrete points from simulations match well with continuous line from Eq. (\ref{eq:mu_star_s}). We use $L=64$ here. \textbf{(b):} The membrane velocity at $\mu^*$ plotted against ${\cal S}/L$ shows a power law increase. The solid line represents a function $\sim x^{0.87}$ and goes parallel to our numerical data points. Here we have used $L=64$. Other simulation parameters are as in Table \ref{table}.}
\label{mu_vs_v_diff_S} \end{figure}

So far we have considered the case for ${\cal S}=L$ membrane updates in one MC step. What happens when the membrane dynamics is faster or slower than this? Apart from the binding site, all the other $(L-1)$ bulk sites are being driven by only the elastic interaction. For large $\cal S$. when the bulk sites are updated at a higher rate, the effect of the elastic interaction is felt more strongly. As a result, the point of balance $\mu^\ast$ where the elastic force and polymerization force become equal, now shifts towards a smaller value of $\mu$. More specifically, for small but nonzero $\mu$, when the bulk sites experience an upward bias towards the binding site, their total displacement in the upward direction per MC step increases, as $\cal S$ increases. Thus the bulk sites are able to catch up with the binding site at a smaller value of $\mu$.
In other words, as $\cal S$ increases, $\mu^\ast$ decreases. In Fig. \ref{mu_vs_v_diff_S}(a) we plot $\mu-V$ data for different ${\cal S}/L$ values, and find that the peak of the curve shifts towards left as ${\cal S}/L $ increases. For ${\cal S}/L >>1 $, the value of $\mu^*$ becomes infinitesimally small and for any finite $\mu$, the curve looks monotonic. In the inset of Fig. \ref{mu_vs_v_diff_S}(a) we show the variation of $\mu^\ast$ with ${\cal S}/L$. Note, however, that the dependence is not very strong and $\mu^\ast$ deceases logarithmically slowly with ${\cal S}/L$. We see that over two decades of change of $\cal S$, the value of $\mu^\ast$ changes only by a factor of half. However, the membrane velocity measured at $\mu^\ast$ shows a more strong dependence on ${\cal S}/L$ and grows as a power law with an exponent $\simeq 0.87$ [Fig. \ref{mu_vs_v_diff_S}(b)]. We can generalize our calculation in the last subsection for arbitrary values of $\cal S$ and obtain $\mu^\ast $ as the only one physical solution of the equation
\be 
\frac{\cal S}{L} \left ( W_0 e^{3 \beta \mu^\ast d} -U_0e^{2 \beta \mu^\ast d} -U_0 e^{ \beta \mu^\ast d} +U_0 \right ) - (U_0^2-U_0 W_0)e^{\beta \mu^\ast d}   =0. 
\label{eq:mu_star_s}
\ee
The analytical result shown by the continuous line in the inset of Fig. \ref{mu_vs_v_diff_S}(a) shows good agreement with numerical data over a wide range of ${\cal S}/L$.

\begin{figure}[h!]
\includegraphics[scale=1.4]{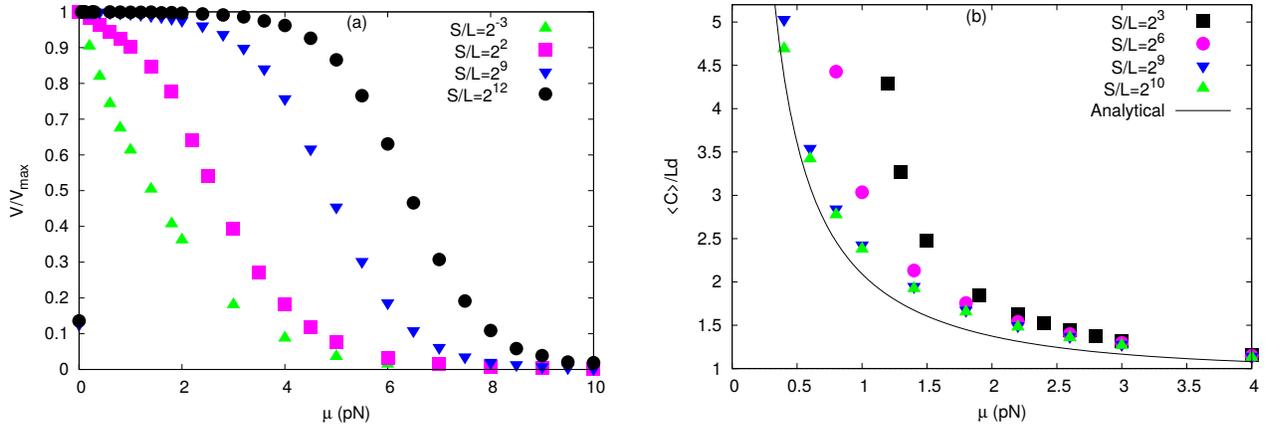}
\caption{ Variation of membrane velocity and contour length against $\mu$ for different values of ${\cal S}/L$. \textbf{(a):} The $\mu-V$ curve changes from convex to concave as ${\cal S}/L$ is increased. For ${\cal S}/L << 1$, the curve is convex while for ${\cal S}/L >> 1$, it becomes concave. Here, we have scaled $V$ by $V_{max}$ such that we can compare them in the same scale. The values of $V_{max}$ are $\num{4.67e-1} nm/s$, $\num{4.67e-1}, nm/s$, $3.73 nm/s$, and $3.73 nm/s$, respectively, in the increasing order of ${\cal S}/L$. Here, we have used $L=8$.  \textbf{(b):} Average contour length of the membrane scaled by $Ld$ as a function of $\mu$ for different values of ${\cal S}/L$. The continuous line is from analytical calculation in the equilibrium limit, which matches with simulation for very high ${\cal S}/L$. Other simulation parameters are as in Table \ref{table}.}
\label{mu_v_large_S}
\end{figure}

Note that our calculation for the contact probability $p_0$ presented in Appendix \ref{app:p0} shows that for small $\mu$ and large ${\cal S}/L$, the contact probability is very small. This means that the filament is unbound most of the time and grows like a free filament with a net growth rate $d(U_0-W_0)$, which is independent of $\mu$. Since we measure the membrane velocity as the average displacement of the membrane per Monte Carlo step and we define our Monte Carlo step such that there are ${\cal S}$ surface updates and one filament update, our data in Fig. \ref{mu_v_large_S}(a) show that the membrane velocity also approaches this limit as ${\cal S}/L$ becomes large. The $\mu -V$ curve in this case therefore becomes flat for small $\mu$ and then decreases for large $\mu$, giving rise to a concave curve, as shown in Fig. \ref{mu_v_large_S}(a). Thus the $\mu -V$ curve can change its shape from convex to concave depending on the choice of the relative time scale ${\cal S}/L$ \cite{Sadhu2016,mogilner2012,carlsson2014}.

In the limit of very large ${\cal S}/L$, when the membrane fluctuations occur much more rapidly than the filament polymerization, the membrane reaches a thermal equilibrium between two filament movements. The partition function for the system can be calculated in this case and the average contour length $\langle {\cal C} \rangle$ of the membrane can be obtained from there, which has the form $\langle {\cal C} \rangle = Ld/(1-e^{-\beta \mu d}) $. 
In Fig. \ref{mu_v_large_S}(b) we compare it with numerics and find reasonably good agreement for large ${\cal S}/L$. 

\subsection{For large $\mu$ the membrane behaves as a rigid obstacle}
\label{sec:rigid}

\begin{figure}[h!]
\includegraphics[scale=1.4]{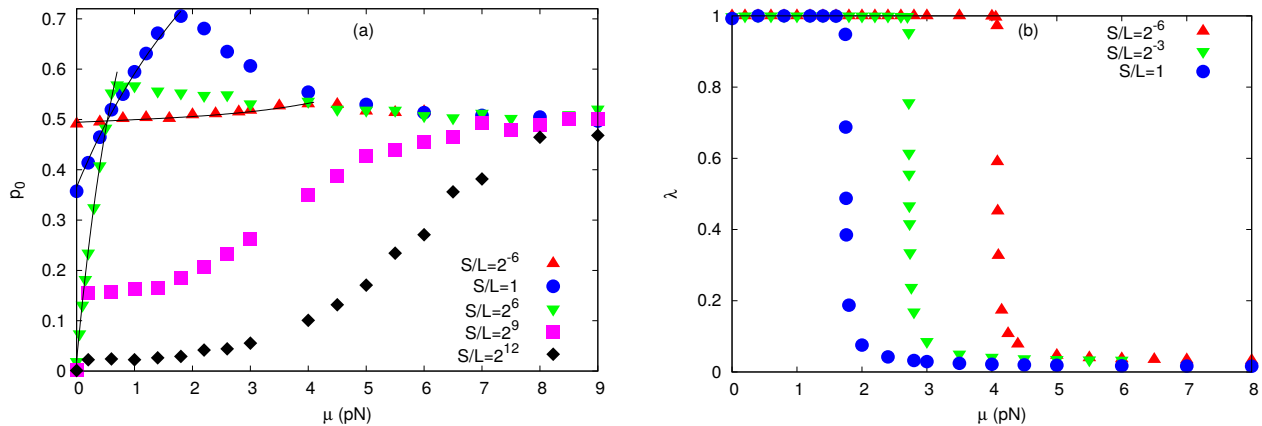}  
\caption{For $\mu >> 1$, the membrane behaves as a rigid obstacle. \textbf{(a):} The contact probability $p_0$ for different values of ${\cal S}/L$. The continuous lines are from the analytical predictions. For large $\mu$, the value of $p_0$ saturates to $1/2$, which is expected for a rigid barrier. As ${\cal S}/L$ decreases, the rigid barrier behavior sets in for smaller $\mu$ values. The continuous lines show analytical calculation in $\mu < \mu^\ast$ regime. \textbf{(b):} The variation of $\lambda$ for different values of ${\cal S}/L$. For small $\mu$, the value of $\lambda$ is unity, which means that the contour length of the membrane diverges with time. For high enough value of $\mu$, $\lambda$ saturates to $2/L$. We use $L=64$ for both the plots. Other simulation parameters are as in Table \ref{table}.}
\label{mu_vs_p0_pi} \end{figure}

When $\mu$ is very large, the membrane remains flat most of the time. Whenever there is a filament polymerization, the height of the  binding site increases, but due to high elastic energy cost associated with such a configuration, the bulk sites quickly catch up and the membrane is flat again. We find that the membrane behaves like a perfectly rigid barrier in this case. Let $p_i$ be the probability to find a gap of length $i$ between the filament tip and the binding site. For a rigid barrier, this gap can only increase or decrease due to filament polymerization or depolymerization. For $i >0$ the master equation for a rigid barrier is 
\be 
\frac{dp_i}{dt}=U_0 p_{i+1}+W_0 p_{i-1}-(U_0+W_0) p_i
\ee
and for $i=0$, 
\be 
\frac{dp_0}{dt}=U_0 p_1-W_0 p_0. 
\ee
In the steady state, we have $p_i=(W_0/U_0)^i p_0 $ and along with normalization condition $\sum_i p_i =1$, this gives the contact probability $p_0 = 1-W_0/U_0 \simeq 1/2$ (see Table \ref{table}). In Fig. \ref{mu_vs_p0_pi}(a), we show the variation of $p_0$ with $\mu$ for different ${\cal S}/L$ and find that for large $\mu$, the contact probability indeed saturates to $1/2$, the rigid barrier limit. Smaller the value of ${\cal S}/L$, faster is the saturation. For small $\mu$, the contact probability can be calculated analytically (see Appendix \ref{app:p0}) and in Fig. \ref{mu_vs_p0_pi}(a) this has been shown by solid lines, which agree well with the simulations. Our calculation remains valid only for $\mu < \mu^\ast$ and hence the comparison has been done only in this regime. $\mu^\ast$ becomes too small for very large ${\cal S}/L$, and hence not marked in this plot. In the case for ${\cal S}/L << 1$, when the membrane dynamics is very slow, the thermal fluctuations of the membrane can be neglected and the contact probability is essentially controlled by the polymerization and depolymerization of the filament. Thus we find that the contact probability approaches $1/2$ even for $\mu < \mu^\ast$ in this case [shown by red triangles in Fig. \ref{mu_vs_p0_pi}(a)]. However, the system is not in steady state here and the contour length of the membrane keeps increasing with time. In Fig.  \ref{mu_vs_p0_pi}(b) we plot the ratio $\lambda = <(|h_{b+1}-h_b| + |h_{b}-h_{b-1}| + 2d )/{\cal C}> $ and show this explicitly. 

\subsection{Results for multiple filaments}
\begin{figure}[h!]
\includegraphics[scale=1.4]{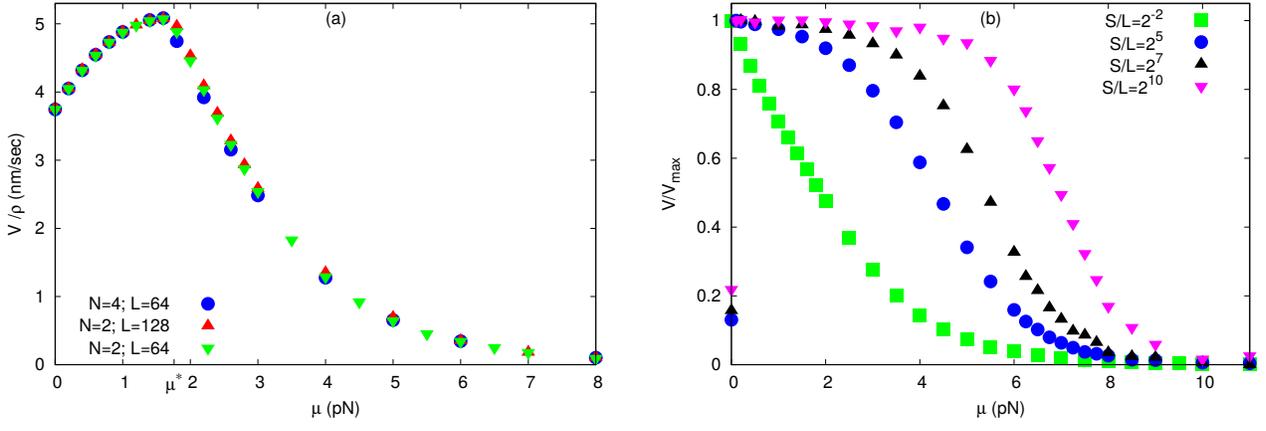}
\caption{$\mu-V$ curve for multiple filaments with $\rho \ll 1$. \textbf{(a):} The velocity scales with $\rho$ similar to the single filament case. It also shows a peak at $\mu=\mu^*$ which is independent of $\rho$. We have used ${\cal S}/L=1$ here. \textbf{(b):} Scaled velocity as a function of $\mu$ for different time scales which show that the curve changes from convex to concave similar to the single filament case. For ${\cal S}/L <<1$, the curve is convex while for ${\cal S}/L >>1$, the curve becomes concave. The values of $V_{max}$ are $0.467 nm/s$, $3.71 nm/s$, $3.74 nm/s$ and $3.75 nm/s$ for ${\cal S}/L=2^{-2}, 2^5, 2^7$, and $2^{10}$, respectively. Here we have used $N=4$ and $L=32$. Other simulation parameters are as in Table \ref{table}.}
\label{mu_vs_v_large_N}
\end{figure}

So far in this section, we have considered the case of single filament. We end this section with a brief discussion on multiple filaments. Let us define the filament density as $\rho =N/L$. We find for $\rho \ll 1$, and for uniform distribution for the binding sites, the qualitative behavior is same as that for single filament. In Fig. \ref{mu_vs_v_large_N}(a), we show the $\mu-V$ curve for multiple filaments for small $\rho$. Since the membrane velocity results from the polymerization force exerted at the binding sites (motion at all other sites follows local detailed balance, as mentioned in Sec. \ref{sec:model}), our data show that $V$ scales as $\rho$. The nonmonotonicity of the $\mu-V$ curve means that even for multiple filaments, the system fails to reach a steady state for small $\mu$ and the membrane contour length $\cal C$ keeps increasing with time. We have checked that (data not shown) at $\mu = \mu^\ast$ the contour length grows diffusively with time and for $\mu > \mu^\ast$ system reaches a steady state when $\cal C$ has a finite value. Figure \ref{mu_vs_v_large_N}(b) shows the $\mu-V$ curve for different values of ${\cal S}/L$. We note that for ${\cal S}/L <<1$, the curve is convex while for ${\cal S}/L >>1$, it becomes concave. 

\begin{figure}[h!]
\includegraphics[scale=0.7]{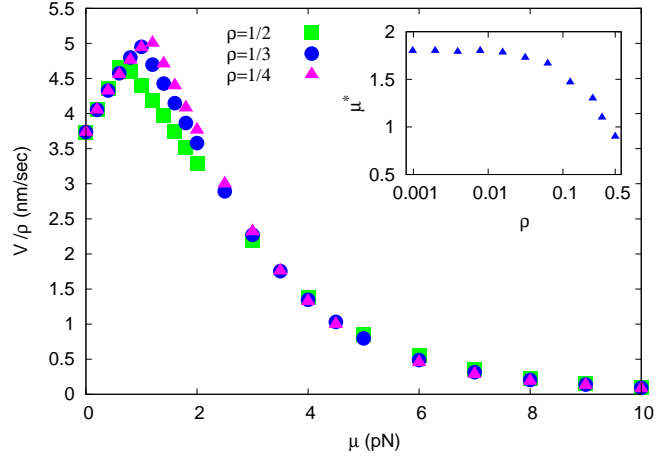}
\caption{Results for multiple filaments with high filament density. In the main plot, we show $\mu-V$ curve for multiple filaments for $\rho \sim 1$. We note that $V$ does not scale with $\rho$. The peak position $\mu^*$ decreases with $\rho$. The inset shows the variation of $\mu^*$ with $\rho$. We note that for $\rho << 1$, the value of $\mu^*$ remains constant and then decreases with $\rho$ for high value of $\rho$. We use ${\cal S}/L=1$ here. Other simulation parameters are as in Table \ref{table}.}
\label{mu_vs_v_high_density}
\end{figure}

As $\rho$ increases, the elasticity of the membrane induces an effective interaction between the filaments and the single filament picture does not remain valid any more. Our data in Fig. \ref{mu_vs_v_high_density} show that $V \sim \rho$ scaling is lost and $\mu^\ast$ now depends on $\rho$. For large $\rho$, most of the sites are binding sites and are driven by the filaments. Due to elastic interaction, the remaining few bulk sites feel an upward pull and are able to catch up with the binding sites at smaller $\mu$ values. Thus $\mu^\ast$ decreases with $\rho$ for large $\rho$. When $\rho =1$, the system will reach steady state for all values of $\mu$ but for $\rho <1$, there will always be a nonzero $\mu^*$ below which the system does not have a steady state. 

In the above discussion, we have assumed that the binding sites are distributed uniformly throughout the membrane. If we consider an inhomogeneous distribution of binding sites, then for small $\mu$ the region of the membrane, where the density of binding sites is high, will have a large local velocity since it is being driven by a large number of filaments. On the other hand, the part of the membrane where binding sites have a low density, will have a much smaller local velocity. In order for the whole system to reach a steady state, the velocity should be same everywhere on the membrane. This means that the polymerization force present in the fastest part of the membrane has to be smaller than the elastic interaction which pushes the slowest part of the membrane where binding site density is lowest. A very large value of $\mu$ is required to achieve this balance. Thus $\mu^\ast$ becomes very large in the case of inhomogeneous distribution of binding sites. We have verified this in our simulations (data not shown here).

\section{Results for the Gaussian model}
\label{sec:gaussian}

In the Gaussian model, the membrane Hamiltonian is given by Eq. (\ref{eq:hgauss}). In this case, if the height of the $j$-th site, $h_j$ changes to $h_j \pm d$, then the change in energy, $\Delta E=2d \nu [d \pm (2h_j-h_{j+1}-h_{j-1})]$, depends on the local height configuration around the $j$th site. This is different from the gradient model, where $\Delta E$ is constant. Following the local detailed balance [see Eq. (\ref{eq:ldb})] we use in our simulations $R_+=e^{-\beta \Delta E}$, and $R_- =1$, while all other movements of the membrane where the energy of the barrier does not change, occur with rate unity. First we present our results for the single filament and then we discuss the case of multiple filaments.

\subsection{Membrane velocity decreases with $\nu$}
\begin{figure}[h!]
\includegraphics[scale=1.4]{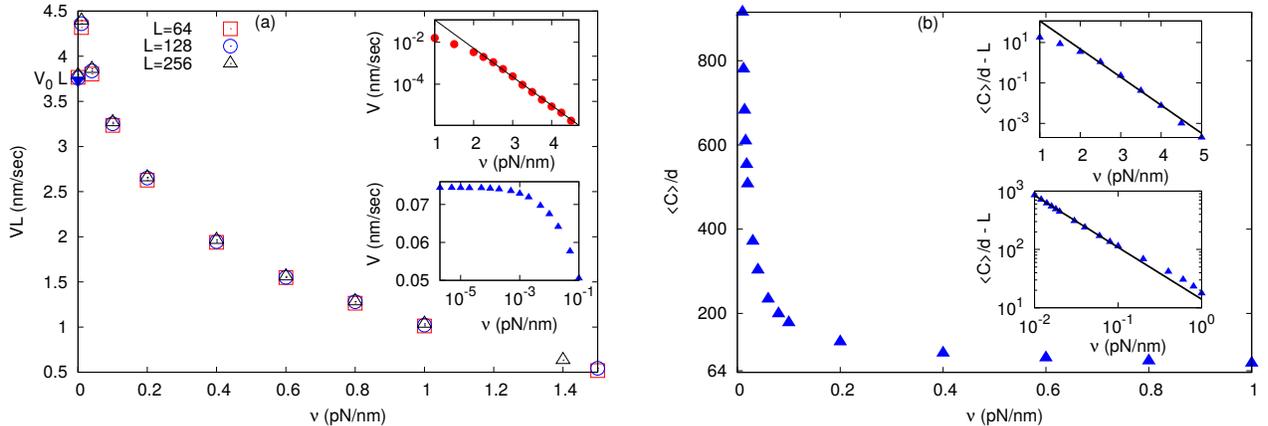} 
\caption{ Variation of velocity and contour length with $\nu$ for Gaussian model. \textbf{(a):} The $\nu-V$ curve for single filament shows $1/L$ scaling. The curve is monotonic for all $\nu \neq 0$. Top inset: The velocity falls exponentially with decay constant $ \simeq 3.13$, which is close to $2 \beta d^2$. Bottom inset: In the limit of very small $\nu$, the velocity saturates to a value $ \sim 0.074 nm/sec$, that is distinctly different from $V_0 = 0.058 nm/s$ at $\nu=0$. We use $L=64$ for the insets and  ${\cal S}/L=1$ for all the plots.
\textbf{(b):} Average contour length of the membrane scaled by $d$ as a function of $\nu$. For very high value of $\nu$, $<{\cal C}>$ becomes equal to the system size $Ld$. Top inset:  Contour length decreases exponentially for large $\mu$ with a decay constant $\simeq 3.2$ which is again close to $2 \beta d^2$. Bottom inset: For very small $\nu$, contour length falls off with $\nu$ as a power law with an exponent $\simeq {0.9}$. For all the above plots, we use $L=64$ and  ${\cal S}/L=1$. Other simulation parameters are as in Table \ref{table}.}
\label{gaussian_mu_vs_v}
\end{figure}

For $\nu=0$, there is no difference between gradient model and Gaussian model [Eqs. (\ref{eq:hamiltonian}) and (\ref{eq:hgauss})]. Therefore, as derived in Sec. \ref{sec:peak}, the membrane velocity at $\nu =0$ is given by $V_0 = d(U_0-W_0)/L$. The membrane contour length $\cal C$ keeps increasing with time and the system does not have a steady state. However, as soon as $\nu$ is nonzero, any polymerization event or independent thermal fluctuation that causes an increase in $\cal C$, has an energy cost which is higher as $\cal C$ gets larger. Thus the membrane can not stretch indefinitely and for all $\nu > 0$ the system reaches a steady state. Our data in Fig. \ref{gaussian_mu_vs_v}(a) show that for $\nu >0$, the membrane velocity $V$ decreases as $\nu$ increases. For large $\nu$, the decay is exponential and we also find $V \sim 1/L$ scaling, as seen earlier in Fig. \ref{mu_vs_v} for the gradient model. 

Note that there is a discontinuity of the $\nu-V$ curve at $\nu =0$ [see Fig. \ref{gaussian_mu_vs_v}(a)]. The data in Fig. \ref{gaussian_mu_vs_v}(a) bottom inset show that as $\nu \to 0$, the velocity saturates to a value, which is different from the value $V_0 = d(U_0-W_0)/L $ at $\nu=0$. This is because at $\nu=0$, only the binding site of the membrane has nonzero velocity and the bulk sites have zero velocity. But as soon as $\nu \neq 0$, the system has a steady state and the bulk sites must move with the same velocity as the binding site. This sharp jump in the bulk sites velocity causes the discontinuity in the $\nu-V$ curve.

In Fig. \ref{gaussian_mu_vs_v}(b) we plot the steady state average contour length $\langle \cal C \rangle $ as a function of $\nu$. We find that (top inset) $\langle \cal C \rangle$ decreases exponentially for large $\nu$, as found in the gradient model (also see Fig. \ref{fig:cl}). For very small $\nu$, our data [Fig. \ref{gaussian_mu_vs_v}(b) bottom inset] show that  $\langle \cal C \rangle $  decreases as a power law with an exponent $\simeq 0.9$. The power law decay can be explained as follows. 

For very small $\nu$, the local height gradient of the membrane decreases sharply as a function of the distance from the binding site and we have $y_b > y_{b+1} > y_{b+2} ...$, where $y_i=h_{i}-h_{i+1}$ and $b$ denotes the binding site. The bound filament polymerization and thermal height fluctuations at the binding site may change $h_b$ which actually changes $ \cal C $. It is easy to see that the rate of these processes that increase  $h_b$ is $ (1+U_0p_0) e^{-2 \beta \nu d^2 (2 y_b+1)}$ and that decrease $h_b$ is ($1-p_0$). The height fluctuations of the other sites will not change $ \cal C $. In the steady state, $\cal C$ is time-stationary, the increasing and decreasing rates must be equal, from which it follows that $ y_b \sim 1 / \nu$. Moreover, in the steady state, the velocity of all membrane sites are equal. The velocity at the binding site is given by $\{ (1+U_0p_0) e^{-2 \beta \nu d^2 (2 y_b+1)} -(1-p_0) \}$, and that at the neighboring site $(b+1)$ is  $\{ 1-e^{-2 \beta \nu d^2 ( y_b -y_{b+1}+1)} \}$. Once we equate them, it follows that $y_{b+1} \sim 1/ \nu$ for small $\nu$. Equating the velocity of the sites $(b+1)$ and $(b+2)$, it can be similarly shown that $y_{b+2} \sim 1/ \nu$, and thus for any site $i$, $y_i \sim 1/\nu$ holds. Thus the quantity ($<{\cal C}> -Ld = \sum_i |y_i| $) shows an $1/ \nu$ dependence. Our numerical data yields power law exponent $0.9$ which is close to this prediction.

\subsection{Faster membrane dynamics yields a concave $\nu-V$ curve}
\begin{figure}[h!]
\includegraphics[scale=1.4]{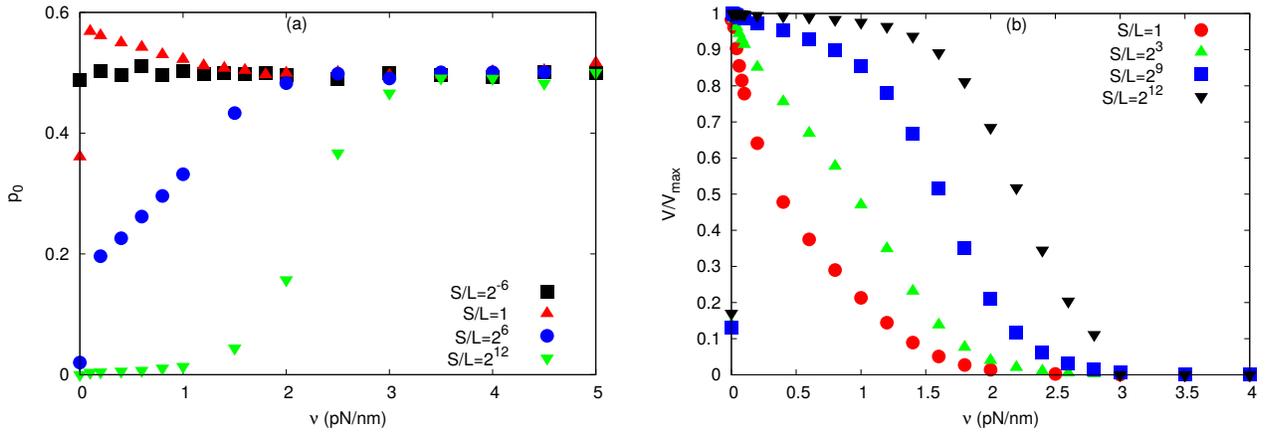} 
\caption{Contact probability and velocity as a function of $\nu$ in a Gaussian model for different values of $ {\cal S}/L$. \textbf{(a):} For ${\cal S}/L <<1$, $p_0$  does not show much variation and its value remains close to $1/2$. For ${\cal S}/L \sim 1$, in the small $\nu$ region, the value of $p_0$ is higher than $1/2$ which then saturates to its rigid barrier limit for high $\nu$. For ${\cal S}/L >>1$, $p_0$ starts with very small value and then increases with $\nu$, then saturates to $1/2$. Higher the value of ${\cal S}/L$, slower the tendency to reach $1/2$. We use $L=8$ here. \textbf{(b):} The $\nu-V$ curve changes from convex to concave as $ {\cal S}/L$ is increased. The values of $V_{max}$ are $0.475 nm/s$, $3.50 nm/s$, $3.72 nm/s$, and $3.73 nm/s$ for ${\cal S}/L=1, 2^3, 2^9$ and $2^{12}$ respectively. Other simulation parameters are as in Table \ref{table}.}
\label{gaussian_mu_vs_p0}
\end{figure}

The relative time-scale between the membrane and filament dynamics is an important parameter for the Gaussian model also. The contact probability $p_0$ that is crucial to determine the interaction between the filament and the membrane depends strongly on whether the membrane dynamics is faster or slower than the filament dynamics. For very large $\nu$, it is expected that $p_0 \simeq 1/2$ since the membrane behaves like a rigid barrier in this limit (also see our discussion in Sec. \ref{sec:rigid}). However, when $\nu$ is small, our data in Fig. \ref{gaussian_mu_vs_p0}(a) show that the variation of $p_0$ with $\nu$ is qualitatively determined by the value of ${\cal S}/L$. For ${\cal S}/L \gg 1$, the thermal fluctuations of the membrane happen so fast that the contact with the filament tip is lost most of the time and $p_0$ is small. From this small value, $p_0$ increases to $1/2$ as $\nu$ becomes large. For ${\cal S}/L \ll 1$, 
the membrane fluctuations become almost negligible and the contact probability is controlled by the (de)polymerization of the (bound) free filament which does not depend on $\nu$. Our data for small ${\cal S}/L$ support this. 

The behavior of $p_0$ has a direct influence on the $\nu-V$ curve and we show in Fig. \ref{gaussian_mu_vs_p0}(b) that the curve becomes concave from convex as ${\cal S}/L$ is increased. When ${\cal S}/L \gg 1$, the filament is most of the time unbound for small $\nu$ and hence the membrane velocity remains constant at $d(U_0-W_0)/L$, same as that for a free filament. For large $\nu$, when $p_0$ starts increasing again, $V$ decreases. The $\nu-V$ curve is concave in this case. On the other hand, for very  small ${\cal S}/L$, the value of $p_0$ is no longer negligible and it remains nearly $0.5$ throughout the region. In this case, due to the increase in $\nu$, the velocity starts decreasing even in the small $\nu$ range, which yields a convex curve.

As done in the gradient model, in the limit of very fast membrane dynamics, we can neglect the filament dynamics and assuming equilibrium we can calculate the average elastic energy of the membrane which shows good agreement with our numerics (data not shown) for very large ${\cal S}/L$.

\subsection{Multiple filaments}
\begin{figure}[h!]
\includegraphics[scale=0.7]{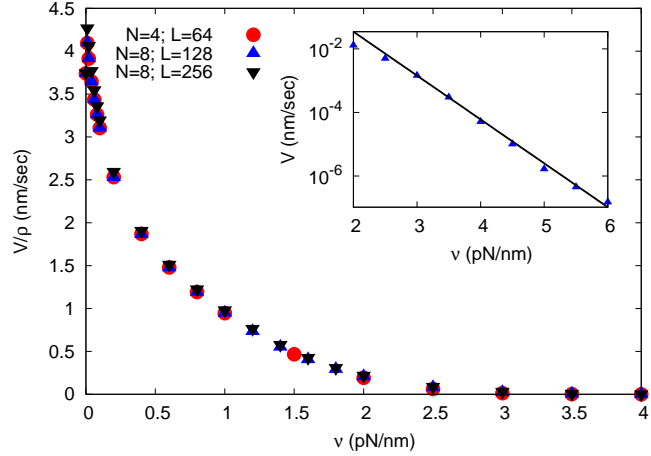} 
\caption{Velocity of the barrier as a function of $\nu$ for multiple filaments. There is no qualitative difference between the results of a single filament and the multiple filaments. For large value of $\nu$, the velocity falls exponentially with a decay constant $ \simeq 3.18$, which is close to the single filament case. For the inset we use $N=8; L=128$ and for all the plots, we use ${\cal S}/L=1$. Other simulation parameters are as in Table \ref{table}.}
\label{gaussian_nu_vs_multiple}
\end{figure}

The multiple filaments case is qualitatively the same as the single filament case. For low density of filaments ($N/L = \rho \ll 1 $), we plot the $\nu-V$ curve in Fig. \ref{gaussian_nu_vs_multiple}. We find that $V$ decreases monotonically with $\nu$ for nonzero $\nu$ and scales as $\rho$. The decay of $V$ is exponential for large $\nu$ with a decay constant which is same as that in the single filament case. In the limit of large $\nu$, the membrane behaves like a rigid barrier and apart from an overall scaling of $V$ by a factor $\rho$, its dependence on $\nu$ remains same for single or multiple filaments.  

\begin{figure}[h!]
\includegraphics[scale=0.7]{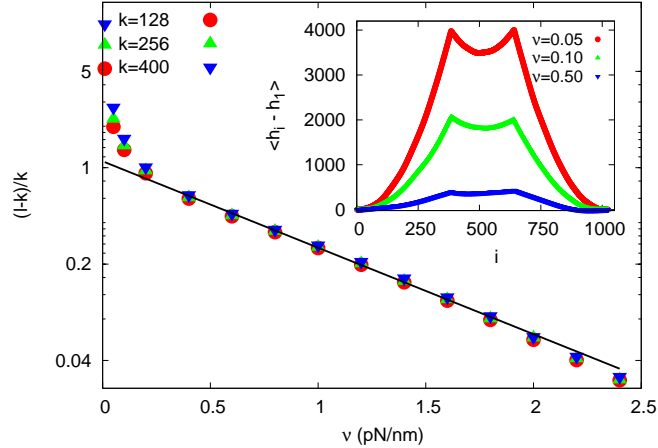} 
\caption{ Merging of protrusions. We plot the quantity $\frac{(l-k)}{k}$, for the membrane with only two binding sites at a distance $k$ apart, with $l$ being the contour length between the two binding sites. For a given $k$, this quantity decreases with $\nu$ exponentially with a decay constant $ \simeq 1.44$, which indicates merging of two protrusions for large $\nu$. Here we use $L=1024$ and ${\cal S}/L=1$. \textbf{Inset:} Two protrusions merge as we increase $\nu$. The average height profile of the membrane is shown for different values of $\nu$. Note that the height of the protrusions also become shorter as $\nu$ increases. The $\nu$ values appearing in the legends are in the unit of $pN/nm$. These data are for $k=256$. We use $L=1024$ and ${\cal S}/L=1$. Other simulation parameters are as in Table \ref{table}.}
\label{merging_protrusion}
\end{figure}

One interesting observation that can be made for the multiple filaments case is the merging of membrane protrusions. In earlier studies \cite{ErdincAtilgan2006} of a more detailed modeling of cell membrane, it was shown that actin filament polymerization gives rise to filopodial protrusions. The protrusion speed depends on elastic properties of the membrane and the membrane distortion also induces an effective attraction between the filopodia and close by filopodia merge together to form a larger (wider) protrusions. We study merging of membrane protrusion within our simple model by monitoring the contour length $l$ of the membrane between two binding sites at a distance $k$ apart. When the protrusions created at the two binding sites merge with each other, $l$ becomes equal to $k$. In Fig. \ref{merging_protrusion}, we plot  $\frac{(l - k)}{k}$  as a function of $\nu$ for a fixed $k$ for the barrier with only two filaments. We find that as $\nu$ is increased, the ratio decreases and the decrease is exponential for moderate to large $\nu$ values. In Fig. \ref{merging_protrusion} inset, we show how the two protrusions merge as $\nu$ increases.

\section{Conclusions}
\label{sec:sum}

In this paper, we study the force generation mechanism by a set of parallel filaments growing against an elastic membrane. The elastic membrane tends to stay flat and any distortion from this flat state costs energy. In our gradient model, where the energy cost is proportional to the absolute magnitude of the local height gradient, we find that the polymerization force wins over the elastic force of the membrane for low values of the membrane tension and the system fails to reach a steady state. This gives rise to a nonmonotonic dependence of membrane velocity on its elasticity. In Gaussian model, where the energy cost of deformation is proportional to the square of the local height gradient, the membrane velocity monotonically decreases with elasticity and the system always reaches a steady state for all nonzero elasticity. Our detailed numerical simulations of various different quantities, accompanied by analytical calculations and simple scaling arguments provide a comprehensive picture of the behavior of the system in the long time limit.

A similar system was studied in Ref. \cite{baumgaertner2010} where the membrane was described by the gradient model and the growing filaments were represented as a network mesh. The filaments always stay below the membrane and can grow only when there is some gap between the mesh and the membrane. This is in contrast to our model where the filaments can push against the membrane and deform it to  make space for themselves to grow. This microscopic difference actually gives rise to qualitatively different scaling behavior for the two systems. We address this important issue in details elsewhere \cite{prep}.

We also highlight the importance of the relative time scale between the dynamics of the membrane and the filaments to determine the qualitative nature of the dependence of membrane velocity on elasticity. While for fast membrane dynamics, the velocity is a concave function of membrane elasticity, for slow membrane dynamics a convex dependence is observed. Both these types of dependencies are observed in real experiments. A convex force-velocity curve is obtained for actin quoted polystyrene beads in Ref. \cite{marcy2004} or for magnetic colloidal particles pushed by unbranched parallel actin filaments in Refs. \cite{baudry2011, baudry2014}. On the other hand, a concave force-velocity curve was reported for branched actin network in Ref. \cite{theriot2005}. Our simple model can yield both these characteristics for different choices of the relative time scale mentioned above.

Throughout this paper, we have only considered elastic energy of the membrane and neglected its bending rigidity. We have also assumed the membrane to be homogeneous with same elasticity along the whole membrane. However, in many physical systems bending rigidity can be important and there can also be tension gradient along the membrane \cite{FogelsonMogilner2014,Yonatan2014}. Our results for the simple model will pave way for studying more complex models
where above mentioned effects are included.

\section{Acknowledgements} 
We acknowledge useful discussions with M. Barma. 
SC acknowledges financial support from the Science and Engineering Research Board, India (Grant No. EMR/2016/001663). The computational facility used in this work was provided through the Thematic Unit of Excellence on Computational Materials Science, funded by Nanomission, Department of Science and Technology (India). 

\appendix
\renewcommand{\theequation}{A-\arabic{equation}}
\setcounter{equation}{0}
\renewcommand{\thefigure}{A-\arabic{figure}}
\setcounter{figure}{0}

\section{Calculation of $p_0$ for the gradient model with single filament for $\mu < \mu^*$}
\label{app:p0}

In this Appendix, we present an analytical calculation for the contact probability $p_0$ in the gradient model with $\mu < \mu^\ast$ with arbitrary ${\cal S}/L$, in presence of a single filament. Let $p_i$ be the probability that there is a gap of size $i$ between the binding site and the filament tip. Clearly, the contact probability $p_0$ corresponds to $i=0$. 
The master equation for $p_i$ can be written as 

$$\frac{dp_i}{dt}=(U_0+ \frac{{\cal S}}{L} e^{\beta \mu d})p_{i+1}+(W_0+ \frac{{\cal S}}{L} e^{-\beta \mu d})p_{i-1}-(U_0+ \frac{{\cal S}}{L} e^{\beta \mu d}+ \frac{{\cal S}}{L} e^{-\beta \mu d}+W_0)p_i; for ~ i>0$$
and
$$\frac{dp_0}{dt}=(U_0+\frac{{\cal S}}{L} e^{\beta \mu d})p_1-(\frac{{\cal S}}{L} e^{-\beta \mu d}+W_0)p_0; for ~ i=0$$
where all symbols have their usual meaning. Here, we have used the fact that for $\mu < \mu^\ast$, the binding site moves faster than the rest of the system and hence has a larger height than all bulk sites. Although the system is not in steady state for $\mu < \mu^\ast$, the probability $p_i$ still reaches a stationary value. In steady state one has the recursion relation $p_i=(\frac{\frac{{\cal S}}{L} e^{-\beta \mu d}+W_0}{\frac{{\cal S}}{L} e^{\beta \mu d}+U_0})^ip_0$; which by applying normalization condition gives the expression for the contact probability,
\be
p_0=\frac{U_0-W_0+\frac{{\cal S}}{L} e^{\beta \mu d}-\frac{{\cal S}}{L} e^{-\beta \mu d}}{\frac{{\cal S}}{L} e^{\beta \mu d}+U_0}; ~for~ \mu < \mu^*.
\label{eq:p0s}
\ee

Alternatively, the binding site being driven by the growing filament, its velocity must be same as the growth velocity of the filament. The later quantity is simply $ U_0 p_0 e^{-\beta \mu d} + U_0 (1-p_0)  -W_0$, while the velocity of the binding site has the form $U_0 p_0 e^{-\beta \mu d} + \frac{{\cal S}}{L} e^{-\beta \mu d} - \frac{{\cal S}}{L} (1-p_0) e^{\beta \mu d}$. Equating these two gives the same expression for $p_0$ as in Eq. (\ref{eq:p0s}).

\renewcommand{\thefigure}{B-\arabic{figure}}
\setcounter{figure}{0}

\section{Probability that $z \geq 0$ and that $h_{b\pm 1} \geq h_{b\pm 2}$}
\label{app:pz}

\begin{figure}[h!]
\includegraphics[scale=1.4]{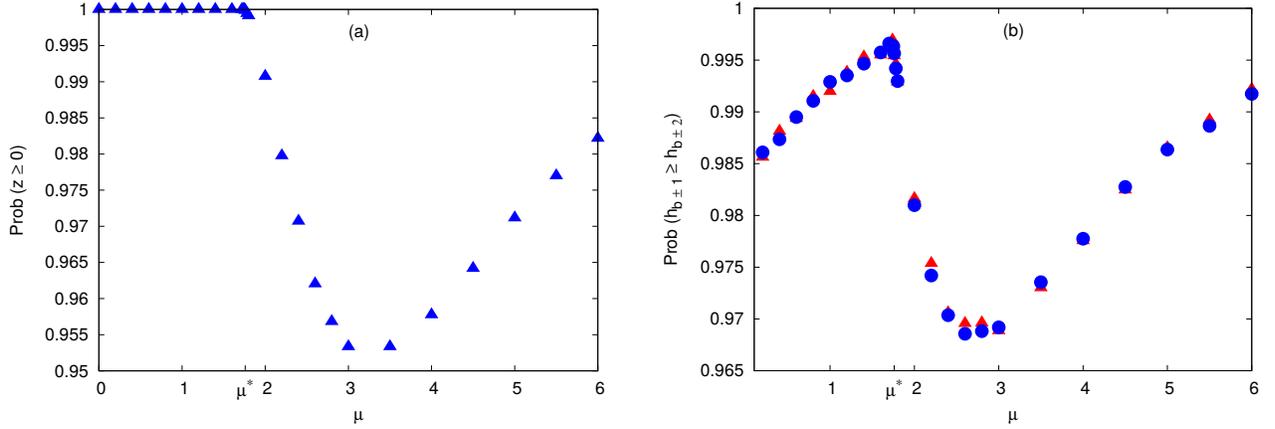} 
\caption{Height gradient around the binding site. \textbf{(a):} The probability that $z \geq 0$  as a function of $\mu$. We see that the probability is exactly one for $\mu < \mu^*$ and even for $\mu > \mu^*$, it is very close to one. \textbf{(b):} Here we show the probability that $h_{b\pm 1} \geq h_{b\pm 2}$ as a function of $\mu$. The red triangles are for $prob (h_{b+1} \geq h_{b+2})$ while the blue circles are for $prob (h_{b-1} \geq h_{b-2})$. These probabilities are also close to one for entire range of $\mu$. Here, $L=64$ and ${\cal S}/L=1$. Other simulation parameters are as in Table \ref{table}.}
\label{height_gradient} \end{figure}

In Fig. \ref{height_gradient}, we plot that the probabilities for  $z \geq 0$ and  $h_{b\pm 1} \geq h_{b\pm 2}$, where $z=2h_b-h_{b+1}-h_{b-1}$. We show that these probabilities are very close to unity even for  $\mu > \mu^*$.



\begin{thebibliography}{99}

\bibitem{review1} J. Howard, \textit{Mechanics of motor proteins and the cytoskeleton}, (Sinauer Associates,  Sunderland, MA, 2001).

\bibitem{review2} T. D. Pollard and J. A. Cooper, Actin, a central player in cell shape and movement, Science {\bf 326}, 1208 (2009).

\bibitem{review3} L. Blanchoin, R. B. Paterski, C. Sykes and J. Plastino, Actin dynamics, architecture and mechanics in cell motility, Physiol Rev {\bf 94}, 235 (2014). 

\bibitem{review4} P. Friedl and D. Gilmour, Collective cell migration in morphogenesis, regeneration and cancer, Nat. Rev. Mol. Cell Biol. {\bf 10}, 445 (2009).

\bibitem{review5} J. Plastino and C. Sykes, The actin slingshot, Curr. Opin. Cell Biol. {\bf 17}, 62 (2005).

\bibitem{review6} P. Sens and J. Plastino, Membrane tension and cytoskeleton organization in cell motility, J. Phys.: Condens. Matter {\bf 27}, 273103 (2015).

\bibitem{review7} A. Diz-Muñoz, D. A. Fletcher and O. D. Weiner, Use the force: Membrane tension as an organizer of cell shape and motility, Trends Cell Biol. {\bf 23}, 47 (2012).

\bibitem{review8} J. Lemière, F. Valentino, C. Campillo and C. Sykes, How cellular membrane properties are affected by the actin cytoskeleton, Biochimie {\bf 130}, 33 (2016).

\bibitem{review9} K. Keren, Cell motility: The integrating role of the plasma membrane, Eur. Biophys. J. {\bf 40}, 1013 (2011).

\bibitem{NilsGauthier2011} N. C. Gauthier, M. A. Fardin, P. Roca-Cusachs, and M. P. Sheetz, Temporary increase in plasma membrane tension coordinates the activation of exocytosis and contraction during cell spreading, Proc. Natl. Acad. Sci. U.S.A. {\bf 108} 14467 (2011).

\bibitem{DRaucher2000} D. Raucher and M. P. Sheetz, Cell spreading and lamellipodial extension rate is regulated by membrane tension, J. Cell Biol. {\bf 148} 127 (2000).

\bibitem{EllenLBatchelder2011} E. L. Batchelder, G. Hollopeter, C. Campillo, X. Mezanges, E. M. Jorgensen, P. Nassoy, P. Sens and J. Plastino, Membrane tension regulates motility by controlling lamellipodium organization,  Proc. Natl. Acad. Sci. U.S.A. {\bf 108}, 11429 (2011).

\bibitem{AllenPLiu2008} A. P. Liu, D. L. Richmond, L. Maibaum, S. Pronk, P. L. Geissler and D. A. Fletcher, Membrane-induced budding of actin filaments, Nat. Phys. {\bf 4}, 789 (2008).

\bibitem{ErdincAtilgan2006} E. Atilgan, D. Wirtz and S. X. Sun, Mechanics and dynamics of actin-driven thin membrane protrusions, Biophys. J. {\bf 90}, 65 (2006).

\bibitem{baumgaertner2010} S. L. Narasimhan and A. Baumgaertner, Dynamics of a driven surface, J. Chem. Phys. {\bf 133}, 034702 (2010).

\bibitem{AVolmer1998} A. Volmer, U. Seifert and R. Lipowsky, Critical behavior of interacting surfaces with tension, Eur. Phys. J. {\bf 5}, 811 (1998).

\bibitem{lipowsky94} R. Lipowsky and S. Grotehans, Renormalization of hydration forces by collective protrusion modes, Biophys. Chem. {\bf 49} 27 (1994); R. Lipowsky and S. Grotehans, Hydration vs. protrusion forces between lipid bilayers, Europhys. Lett. {\bf 23} 599 (1993).

\bibitem{ew} S. F. Edwards and D. R. Wilkinson, The surface statistics of a granular aggregate, Proc. R. Soc. London {\bf 381}, 17 (1982).

\bibitem{kpz} M. Kardar, G. Parisi and Y-C. Zhang, Dynamic scaling of growing interfaces, Phys. Rev. Lett. {\bf 56}, 889 (1986).

\bibitem{baumgaertner2012} A. Baumgaertner, Crawling of a driven adherent membrane, J. Chem. Phys. {\bf 137}, 144906 (2012). 

\bibitem{Sadhu2016} R. K. Sadhu and S. Chatterjee, Actin filaments growing against a barrier with fluctuating shape, Phys. Rev. E {\bf 93}, 062414 (2016).

\bibitem{hansda2014} D. K. Hansda, S. Sen and R. Padinhateeri, Branching influences force-velocity curve and length fluctuations in actin networks, Phys. Rev. E {\bf 90}, 062718 (2014).

\bibitem{carlsson2014} R. Wang and A. E. Carlsson, Load sharing in the growth of bundled biopolymers, New J. Phys {\bf 16}, 113047 (2014).

\bibitem{peskin1993} C. S. Peskin, G. M. Odell and G. F. Oster, Cellular motions
and thermal fluctuations: The Brownian ratchet, Biophys. J. {\bf 65}, 316 (1993).

\bibitem{marcy2004}  Y. Marcy, J. Prost, M. F. Carlier, and C. Sykes, Forces generated during actin-based propulsion: A direct measurement by micromanipulation, Proc. Natl. Acad. Sci. U.S.A. {\bf 101}, 5992 (2004).

\bibitem{baudry2011} C. Brangbour, O. du Roure , E. Helfer, D. Démoulin, A. Mazurier, M. Fermigier, M. F. Carlier, J. Bibette and J. Baudry, Force-velocity measurements of a few growing actin filaments, PLoS Biol. {\bf 9}, e1000613 (2011). 

\bibitem{baudry2014}  D. Démoulin, M. F. Carlier, J. Bibette, and J. Baudry, Power transduction of actin filaments ratcheting in vitro against a load,  Proc. Natl. Acad. Sci. U.S.A. {\bf 111}, 17845 (2014).

\bibitem{theriot2005} S. H. Parekh, O. Chaudhuri, J. A. Theriot and D. A. Fletcher, Loading history determines the velocity of actin-network growth, Nat. Cell Biol. {\bf 7}, 1219 (2005).

\bibitem{kirone} K. Tsekouras, D. Lacoste, K. Mallick and J. F. Joanny, Condensation of actin filaments pushing against a barrier, New J. Phys. {\bf 13}, 103032 (2011).

\bibitem{ddas2014} D. Das, D. Das and R. Padinhateeri, Collective force generated by multiple biofilaments can exceed the sum of forces due to individual ones, New J. Phys. {\bf 16}, 063032 (2014).

\bibitem{nelson} D. Nelson, T. Piran and S. Weinberg, {\it Statistical Mechanics of Membranes and Surfaces}, (World Scientific, Singapore, 2004). 

\bibitem{pollard} T. D. Pollard, Rate constants for the reactions of ATP-and ADP-actin with the ends of actin filaments, J. Cell. Biol. {\bf 103}, 2747 (1986).

\bibitem{krawczyk2011} J. Krawczyk and J. Kierfeld, Stall force of polymerizing microtubules and filament bundles, Europhys. Lett. {\bf 93}, 28006 (2011).

\bibitem{mogilner2012} J. Zhu and A. Mogilner, Mesoscopic model of actin-based
propulsion, PLoS Comp. Biol. {\bf 8}, e1002764 (2012). 

\bibitem{prep} R. K. Sadhu and S. Chatterjee (Unpublished).

\bibitem{FogelsonMogilner2014} B. Fogelson and A. Mogilner, Computational estimates of membrane flow and tension gradient in motile cell, PLoS One {\bf 9}, e84524 (2014).

\bibitem{Yonatan2014} Y. Schweitzer, A. D. Lieber, K. Keren and M. M. Kozlov, Theoretical analysis of membrane tension in moving cells, Biophys. J. {\bf 106} 84 (2014).

\end{thebibliography}
\end{document}